\documentclass[a4p,12pt]{article}
\usepackage[]{cite}
\usepackage[]{epsfig}
\usepackage[]{rotating}
\usepackage[]{time}

\newcommand{\Date}      {22 December, 2000}

\newcommand{\SM}{Standard Model}

\newcommand{\ee}{\mbox{${\mathrm{e}}^+ {\mathrm{e}}^-$}}
\newcommand{\tautau}{\mbox{$\tau^+\tau^-$}}
\newcommand{\mm}{\mbox{$\mu^+\mu^-$}}

\newcommand{\qq}         {\mbox{$\mathrm{q}\bar{\mathrm{q}}$}}
\newcommand{\bb}         {\mbox{$\mathrm{b}\bar{\mathrm{b}}$}}
\newcommand{\ff}         {\mbox{$\mathrm{f}\bar{\mathrm{f}}$}}
\newcommand{\nunu}       {\mbox{$\nu\bar{\nu}$}}
\newcommand{\mZ}         {\mbox{$m_{\mathrm{Z}}$}}
\newcommand{\mH}         {\mbox{$m_{\mathrm{H}}$}}

\newcommand {\Ho}        {\mbox{$\mathrm{H}^{0}$}}

\newcommand {\Zo}        {\mbox{$\mathrm{Z}^{0}$}}

\newcommand{\Zgs}        {\mbox{$\mathrm{(Z/\gamma)}^{*}$}}

\newcommand{\nn}{\mbox{$\nu \bar{\nu}$}}

\newcommand{\WW}         {\mbox{$\mathrm{W}^+\mathrm{W}^-$}}
\newcommand{\pb}         {\mbox{$\mathrm{pb}^{-1}$}}

\newcommand{\sqrts}     {\sqrt{s}}
\newcommand{\PhysLettB}[1] {Phys. Lett. {\bf B#1}}
\newcommand{\CPC}[1]      {Comp.\ Phys.\ Comm.\ {\bf #1}}
\def\etal{\mbox{{\it et al.}}}

\newcommand{\ra}        {\mbox{$\rightarrow$}}   
\begin{document}
\begin{titlepage}
\centerline{\LARGE European Organisation for Nuclear Research}
\bigskip
\begin{flushright}
      CERN-EP/2000-156 \\ 
      \Date
\end{flushright}

\begin{center}{\LARGE\bf Search for the Standard Model Higgs Boson \\
in {\boldmath \ee\unboldmath} Collisions at 
{\boldmath $\sqrt{s} \approx$ \unboldmath}192--209~GeV}
\end{center}
 \bigskip
\begin{center}
{\LARGE The OPAL Collaboration}

\bigskip
\bigskip



\end{center}
\bigskip
\begin{center}{\large  Abstract}\end{center}
A search for the Standard Model Higgs boson has been performed 
with the OPAL detector at LEP based on the full data sample
collected at $\sqrt{s}\approx$~192--209~GeV in 1999 and 2000,
corresponding to an integrated luminosity
of approximately 426 pb$^{-1}$.
The data are examined for their consistency with
the background-only hypothesis and various Higgs boson mass
hypotheses.   A lower bound of 109.7~GeV is obtained on the
Higgs boson mass at the 95\% confidence level.
At higher masses, the data are consistent with both 
the background and the signal-plus-background hypotheses.

\bigskip\bigskip\bigskip\bigskip

\begin{center}
  {\large (Accepted by Physics Letters B)}
\end{center}
\bigskip\bigskip
This work is dedicated to the memory of our friends and collaborators,
Professors Shuji Orito and George~A.~Snow.


\end{titlepage}

\begin{center}{\Large        The OPAL Collaboration
}\end{center}\bigskip
\begin{center}{
G.\thinspace Abbiendi$^{  2}$,
C.\thinspace Ainsley$^{  5}$,
P.F.\thinspace {\AA}kesson$^{  3}$,
G.\thinspace Alexander$^{ 22}$,
J.\thinspace Allison$^{ 16}$,
G.\thinspace Anagnostou$^{  1}$,
K.J.\thinspace Anderson$^{  9}$,
S.\thinspace Arcelli$^{ 17}$,
S.\thinspace Asai$^{ 23}$,
D.\thinspace Axen$^{ 27}$
% ...
}\end{center}\bigskip
\bigskip
$^{  1}$School of Physics and Astronomy, University of Birmingham,
Birmingham B15 2TT, UK
\newline
$^{  2}$Dipartimento di Fisica dell' Universit\`a di Bologna and INFN,
I-40126 Bologna, Italy
\newline
$^{  3}$Physikalisches Institut, Universit\"at Bonn,
D-53115 Bonn, Germany
\newline
% ...
$^{  5}$Cavendish Laboratory, Cambridge CB3 0HE, UK
\newline
% ...
$^{  9}$Enrico Fermi Institute and Department of Physics,
University of Chicago, Chicago IL 60637, USA
\newline
% ...
$^{ 16}$Department of Physics, Schuster Laboratory, The University,
Manchester M13 9PL, UK
\newline
$^{ 17}$Department of Physics, University of Maryland,
College Park, MD 20742, USA
\newline
% ...
$^{ 22}$Department of Physics and Astronomy, Tel Aviv University,
Tel Aviv 69978, Israel
\newline
$^{ 23}$International Centre for Elementary Particle Physics and
Department of Physics, University of Tokyo, Tokyo 113-0033, and
Kobe University, Kobe 657-8501, Japan
\newline
% ...
$^{ 27}$University of British Columbia, Department of Physics,
Vancouver BC V6T 1Z1, Canada
\newline
% ...
Tel Aviv 69978, Israel.

\newpage
\widowpenalty=1000
\section{Introduction}
In the Standard Model~\cite{sm}, the Higgs mechanism~\cite{higgs} gives
mass to the electroweak gauge bosons, thus allowing
the unification of the electromagnetic and weak interactions.
Whether the Higgs boson exists is one of the most important
open questions in particle physics.
A number of improvements were made to the LEP collider for the year 2000
data taking, which increased the Higgs discovery potential, extending the
sensitivity of the Higgs boson search to approximately 115 GeV if the data
from all four LEP experiments are combined.
The combination of the preliminary Higgs boson search
results of the four LEP 
experiments~\cite{higgs_piktalk,higgs_otherlep} 
shows an excess of candidates which may indicate the production of
a Standard Model Higgs boson with a mass near 115 GeV.
In this letter we present the results of a search
for the Standard Model Higgs boson with
the OPAL detector at LEP, considering particularly the mass
hypothesis of 115 GeV.

Approximately 426~\pb\ of e$^+$e$^-$ annihilation data were collected
by OPAL in the years 1999 and 2000 at centre-of-mass
energies in the range 192--209 GeV; this data sample is used for the
analyses presented in this letter.
Searches are performed for 
the ``Higgs-strahlung'' process
\ee\ra\Ho\Zo\ra\Ho\ff, where \Ho\  is 
the \SM\ Higgs boson,
and
\ff\ is a fermion-antifermion pair from the \Zo\ decay.
For the \Ho\nunu\ (\Ho\ee) final state, the contribution from 
the $\WW$ ($\Zo\Zo$) fusion process
is also taken into account.
Only the decays of the Higgs boson 
into \bb\  and \tautau\ are considered here. 
OPAL has already reported results from the \SM\ Higgs boson search at 
e$^+$e$^-$ centre-of-mass energies up to 189 GeV~\cite{pr189,prold}, where
a lower mass limit of $\mH>91.0$~GeV was obtained at the 
$95\%$ confidence level. 
A similar search procedure is applied here. 
All data were
processed with the most up-to-date detector calibrations available. For
future publications more refined analyses and 
the final detector calibrations will be used.

\section{OPAL Detector, Data and Monte Carlo Samples}\label{sect:detector}

Details of the OPAL detector can be found in~\cite{detector}.
The data used in the analyses correspond to integrated luminosities 
of approximately 216 pb$^{-1}$ at 192--202 GeV, 80 pb$^{-1}$
at 203--206 GeV, and 130 pb$^{-1}$ at centre-of-mass energies 
higher than 206 GeV.  The total luminosity used to search for
the Higgs boson varies by $\pm$2\% from channel to channel,
due to slightly different requirements
on the operational status of different detector elements.
During 2000 (1999), data were taken at
$\sqrt{s}=200$-209 GeV (192-202 GeV) with a luminosity-weighted mean
centre-of-mass energy of 206.1 (197.6) GeV. 

A variety of Monte Carlo samples was generated 
at centre-of-mass energies between 192 and 210 GeV.
Higgs boson production is modelled with the HZHA3 generator~\cite{HZHA3}
for a wide range of Higgs boson masses.  The size of these samples
varies from 2000 to 10000 events for each mass and at each centre-of-mass energy.
The background processes are simulated with typically more than 50 times the
statistics of the corresponding data sample.
The process \Zgs\ra\qq($\gamma$) is modelled with
the KK2f generator using CEEX~\cite{CEEX} 
for the modelling of the initial state radiation.  
The four-fermion processes (4f) are simulated using
grc4f~\cite{grc4f}.
The two-photon and other two-fermion processes
have a negligible impact on the results.
The hadronisation process is simulated with 
JETSET/PYTHIA~\cite{pythia} with parameters
described in~\cite{opaltune}.
In each search channel, the estimates of the signal efficiency and 
the selected background rate depend strongly on the centre-of-mass energy 
and are thus interpolated on a fine grid.
For each Monte Carlo sample, the full detector response is simulated 
in detail as described in~\cite{gopal}.

\section{Analysis Procedures}
\label{sect:analysisprocedures}

We search for  
Higgs production
in the following final
states: \Ho\Zo\ra\bb\qq\  (four-jet channel), \Ho\Zo\ra\bb\nn\ 
(missing-energy channel),
\Ho\Zo\ra\bb\tautau\ and \tautau\qq\
(tau channels), \Ho\Zo\ra\bb\ee\ and \bb\mm\ (electron and
muon channels).  In each channel, 
a preselection is applied to ensure that the events are well measured
and are consistent with the desired signal topology. 
A likelihood selection combining 6 to 10
variables depending on the search channel
is then used to further enrich the signal.

\begin{figure}[htb]
\vspace*{-0.2cm}
\centerline{\epsfig{file=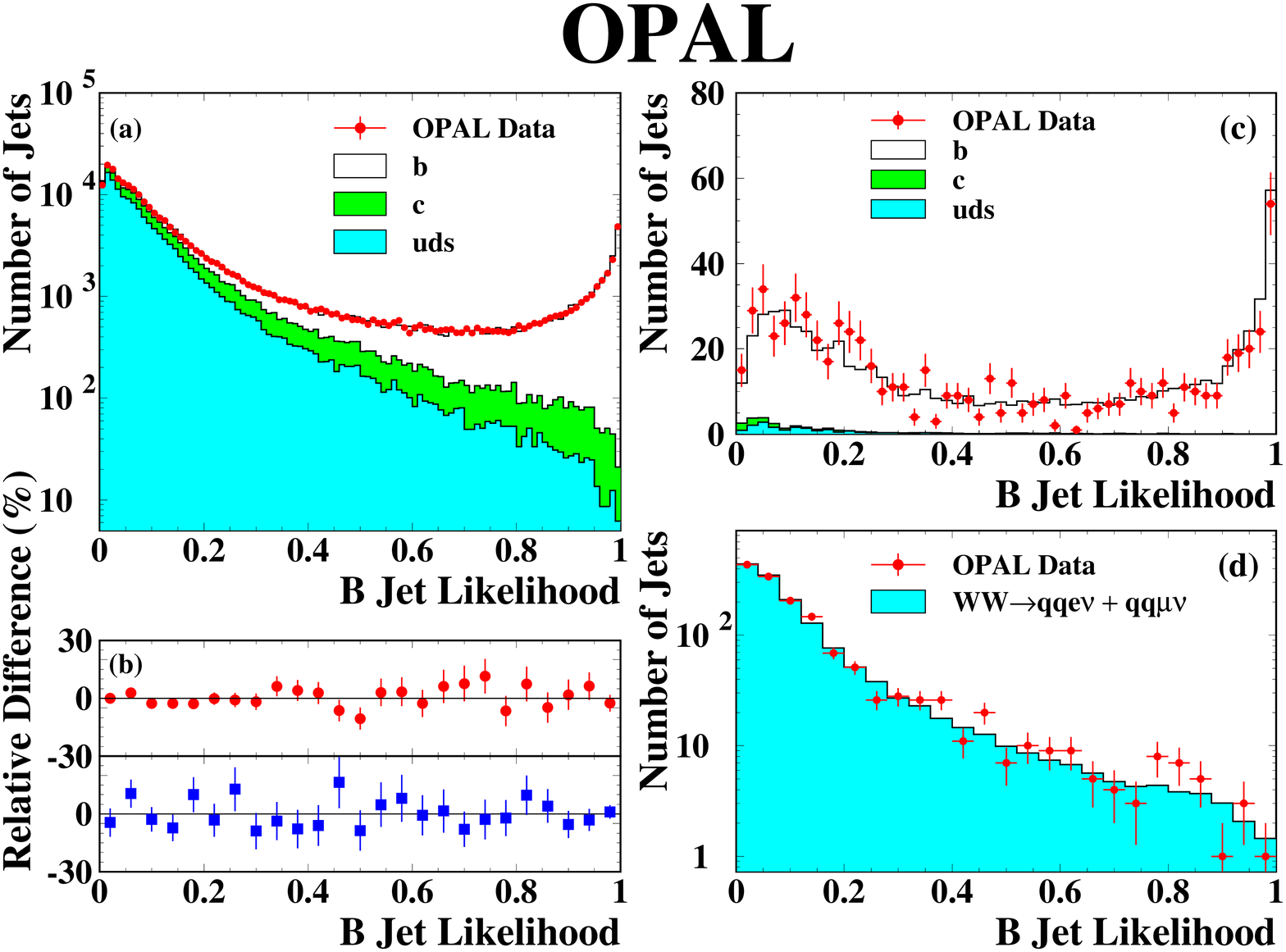,width=1.2\textwidth}}
\vspace{-3cm}
\caption[]{\label{fig:btag}\sl
The b-tagging performance and modelling for (a--b) calibration data 
taken at $\sqrt{s}=\mZ$ in 2000, and (c--d) at $\sqrt{s}$
between 200--209 GeV in 2000.
(a) The distribution of the b-tagging variable $\cal B$ for jets
in data compared to the Monte Carlo expectation. 
(b) The bin-by-bin difference between data and Monte Carlo simulation
for jets opposite non b-tagged jets (circles) and
for jets opposite b-tagged jets (squares). 
(c) The b-tagging output, $\cal{B}$, for jets opposite b-tagged jets
in a sample of $\qq\gamma$ events,
and (d) for jets in a sample of 
${\mathrm {W^{+}W^{-}\rightarrow q{\overline q}}
{\mathrm e^-}{\overline\nu_{\mathrm{e}}}}$ and
${\mathrm {W^{+}W^{-}\rightarrow q{\overline q}}
\mu^-{\overline\nu_{\mu}}}$ events (and charge conjugates).
The  histogram in (d) shows the distribution from the four-fermion
Monte Carlo samples.
}
\end{figure}

We use the same analysis techniques described in a 
previous publication~\cite{pr189}, namely
event reconstruction, b-flavour tagging, lepton (electron,
muon and tau) identification and kinematic fits to reconstruct the Higgs
boson mass. 
 The b-tagging variable ${\cal B}$ is evaluated for each jet
 to distinguish jets containing b-hadrons from those that do not.
  The tracking and b-tagging performance in the Monte Carlo simulation 
  are tuned using 
  8.2~pb$^{-1}$ of calibration data 
  collected at $\sqrt{s}\approx\mZ$ at intervals during 1999 and 2000 
  with the same detector
  configuration and operating conditions as the high-energy data.
  Figure~\ref{fig:btag}(a) shows the distribution of
  ${\cal B}$ for the calibration data. 
  Comparisons between the data and the Monte Carlo simulation are
  shown in Figure~\ref{fig:btag}(b) for
  jets which are found opposite to jets passing or failing
  the b-tagging requirement.
  The tagging efficiency for b-flavoured (udsc-flavoured) jets 
  is modelled by the Monte Carlo simulation to within an 
  accuracy of 2\% (5\%).

The performance of the b-tagging 
for the data taken at $\sqrt{s}\geq 192$~GeV
is checked with samples of
$\qq(\gamma)$ events by selecting hadronic 
events with the mass of the $\qq$ system near \mZ.
Figure~\ref{fig:btag}(c) shows the b-tagging variable $\cal{B}$ 
for jets opposite b-tagged jets in the 2000 sample.
The efficiency for tagging udsc flavours is also
checked by computing ${\cal B}$ for the jets in a sample of
${\mathrm {W^{+}W^{-}\rightarrow q{\overline q}}\ell{\overline \nu}}$ 
($\ell$=e or $\mu$)
obtained with the selection
used to measure the \WW\ cross-section~\cite{ref:wwxs}
as shown in Figure~\ref{fig:btag}(d).
The expectation from the \SM\ Monte Carlo describes the data
within the relative statistical uncertainty of 5--10\%.
The results of these cross-checks are not used
in the evaluation of the systematic uncertainties described below.

\label{sec:syserror}

Sources of systematic uncertainties are investigated for their effect
on the signal detection efficiencies and the Standard 
Model backgrounds.
The error from the modelling of the likelihood selection input variables
on the background (signal) rates
is 4--8\% (1--4\%), depending on the channel.  These uncertainties are
evaluated based on comparisons of the distributions of the variables
in the data and the Monte Carlo.
Comparisons of alternative Monte Carlo generators 
for the backgrounds~\cite{CEEX,grc4f,pythia,koralW}
account for an additional 3--11\% uncertainty in the background rates.
The uncertainty in the four-fermion cross-section is taken
to be $\pm 2$\%~\cite{ref:4funcertainty}.
The uncertainties on the detector performance, such as
the spatial resolution of the tracking and the modelling of the efficiency of the silicon
microvertex detector, are evaluated source by source with Monte Carlo studies.
Recent improvements in the knowledge of 
heavy quark production processes and decays, such as 
the b-hadron charged decay multiplicity~\cite{LEPEWWG} and
the gluon splitting rate to heavy quarks~\cite{gqq-spl}, 
are taken into account in the
analyses by reweighting Monte Carlo events.
The modelling of the b-hadron production and decay processes is constrained by
various measurements summarized 
in~\cite{LEPEWWG}; residual uncertainties in these measurements result
in systematic uncertainties here.
In particular, the b-hadron charged decay multiplicity $n_B$
is varied within $n_B = 4.955\pm 0.062$.
The uncertainties from the fragmentation functions 
for b- and c-quarks are obtained by adjusting the mean 
energy $\langle x_E\rangle$ within the range allowed by the measurements.
The charm and bottom-flavoured hadron lifetimes are varied 
within their
errors with negligible effect on all search channels.
The total uncertainties on the background (signal) rates
are 
11--15\% (5--6\%) varying from channel
to channel.

\subsection{\label{sect:sm4jet} 
Event Selections and Mass Reconstruction}

The preselection requirements in the four-jet channel
are identical to those of~\cite{pr189}.
After the preselection, a likelihood selection based on eight variables
is applied.
For each selected candidate,
two of the jets are associated to the $\Ho$
using a likelihood method based on
the kinematic fit result and the b-tagging information. 
The mass determined by a 
${\mathrm{H^{0}Z^{0}}}$ 5C kinematic fit for the chosen jet pair gives 
the reconstructed mass of the Higgs boson candidate, 
$m^{\mathrm{rec}}_{\mathrm{H}}$.
Because of the 
constraints of the kinematic fit, 
$m^{\mathrm{rec}}_{\mathrm{H}}<\sqrt{s}-\mZ$.
The variables used in the likelihood selection 
for the 1999 data are the same as those used in our earlier analysis~\cite{pr189}.
For the 2000 data, however, one variable, the $\chi^{2}$ probability 
of the ${\mathrm{H^{0}Z^{0}}}$ 5C kinematic fit,
is replaced by the $\chi^{2}$ probability of a fit imposing an equal-mass constraint
as used in the OPAL W mass measurement~\cite{WWpaper}
in order to further suppress the $\Zo\Zo$ background.

The selection of the missing-energy channel is very similar
to~\cite{pr189}.
The notable changes are:
1) tightening the requirement on the maximum
fraction of the visible energy
in the angular region $|\cos{\theta}| > 0.90$ from 50\% to 20\%
in order to further suppress $\qq(\gamma)$ background 
(changed only in the 2000 analysis);
2) a looser requirement on the missing mass  
which is now selected in the range from 40 to 140~GeV.
Other small changes correspond to the scaling of cut values with $\sqrt{s}$
for variables related to the visible momenta.
In the construction of the likelihood selection,
two new variables are included:
1) the thrust value of the event, and 2) 
the angle between the missing momentum 
and the direction of the most energetic jet.
This last variable adds discrimination power especially against
the process 
$\mathrm {W^{+}W^{-}\rightarrow q{\overline q}}\ell{\overline \nu}$
in which
the charged lepton is close to one of the jet axes.
The reconstructed Higgs boson mass $m^{\mathrm{rec}}_{\mathrm{H}}$
is evaluated using a kinematic fit constraining
the missing mass to the \Zo\ mass. 
Because of the fit constraints,
$m^{\mathrm{rec}}_{\mathrm{H}}<\sqrt{s}-\mZ$.

The event selections in the tau, electron and muon channels are
identical to the ones used in \cite{pr189}.
For the tau channel the reconstructed
Higgs boson mass is evaluated (see \cite{pr189}) with the 3C 
kinematic fit  
with the largest $\chi^2$ probability
fixing either the
tau pair invariant mass or the jet pair 
invariant mass to the \Zo\ mass.
The reconstructed Higgs boson mass 
is determined by the recoil mass of the
electron pair in the electron channel, and with the results of a
4C kinematic fit constraining energy and momentum in the muon channel.
There is no upper bound of $m^{\mathrm{rec}}_{\mathrm{H}}$ at
$\sqrt{s}-\mZ$ in the electron and muon channels.

\begin{figure}[ht]
\vspace*{-0.7cm}
\centerline{
\epsfig{file=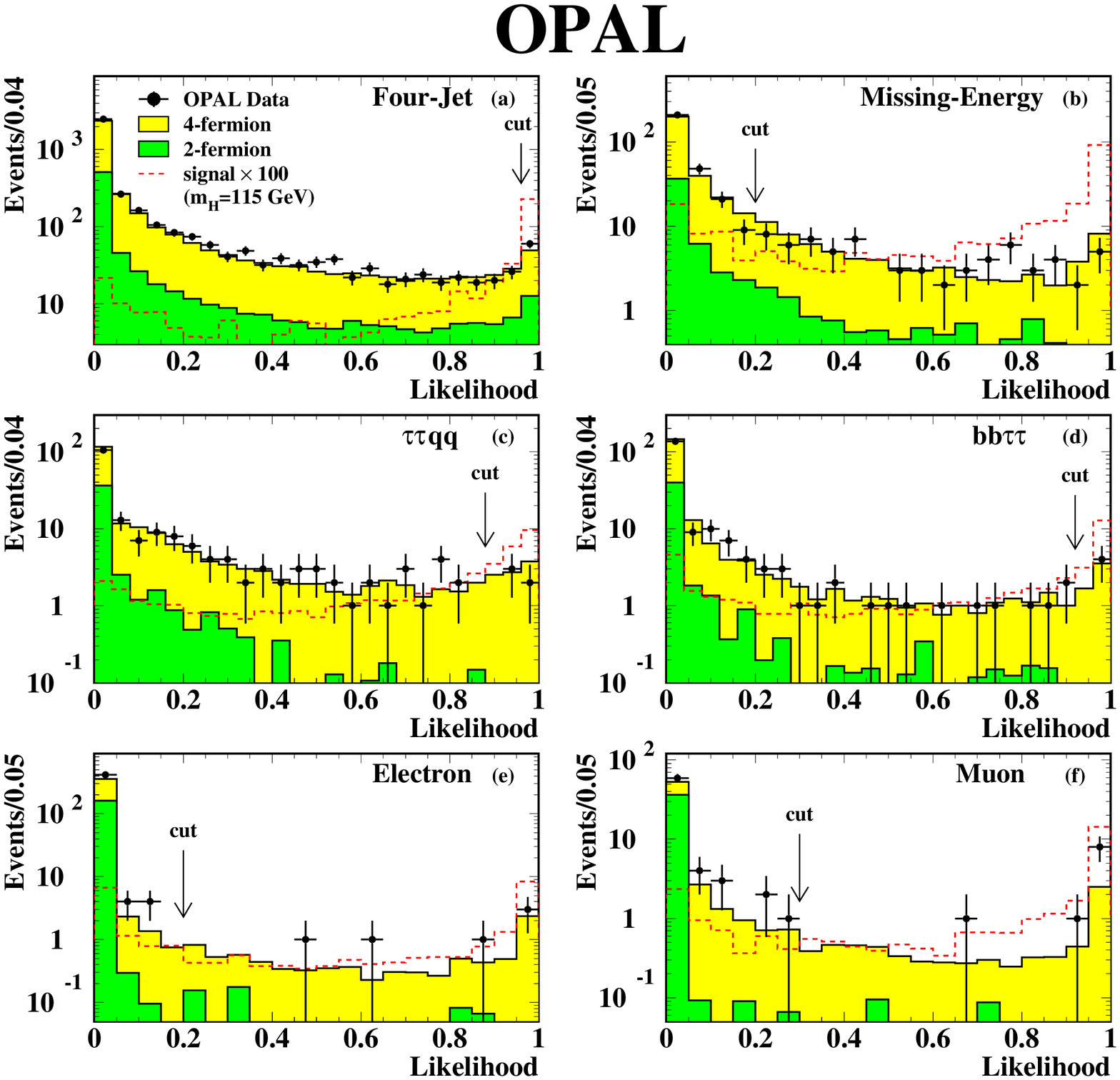,width=1.2\textwidth}
}
\vspace*{-1cm}
\caption[]{\label{fig:likeout}\sl
The distributions of the likelihood output variables in \Ho\Zo\ searches for
the 1999 and 2000 data in:
(a) the four-jet channel, (b) the missing-energy channel,
(c) the tau channel $\tautau\qq$, (d) the tau channel $\bb$\tautau,
(e) the electron channel and (f) the muon channel.  
OPAL data are shown with points, 
backgrounds with the shaded histograms, 
and
the expectation from a signal with \mH=115~GeV with the dashed histograms
(scaled up by a factor of 100 for visibility.)
}
\end{figure}

The numbers of events selected in each analysis 
after preselection
and after the final likelihood selection are shown
in Table~\ref{tab:tab1new} for the data taken at $\sqrt{s}\approx 192-209$ GeV.
The errors on the background and signal expectations 
are the sums in quadrature of the individual systematic uncertainties.
Distributions of the selection likelihood values ${\cal L}^{\rm H^0Z^0}$
in all channels are shown in Figure~\ref{fig:likeout}.
The number of selected events in all search channels 
is 156 with 146.1$\pm$11.9 expected from \SM\ background
processes.

\begin{figure}[ht]
\centerline{
\epsfig{file=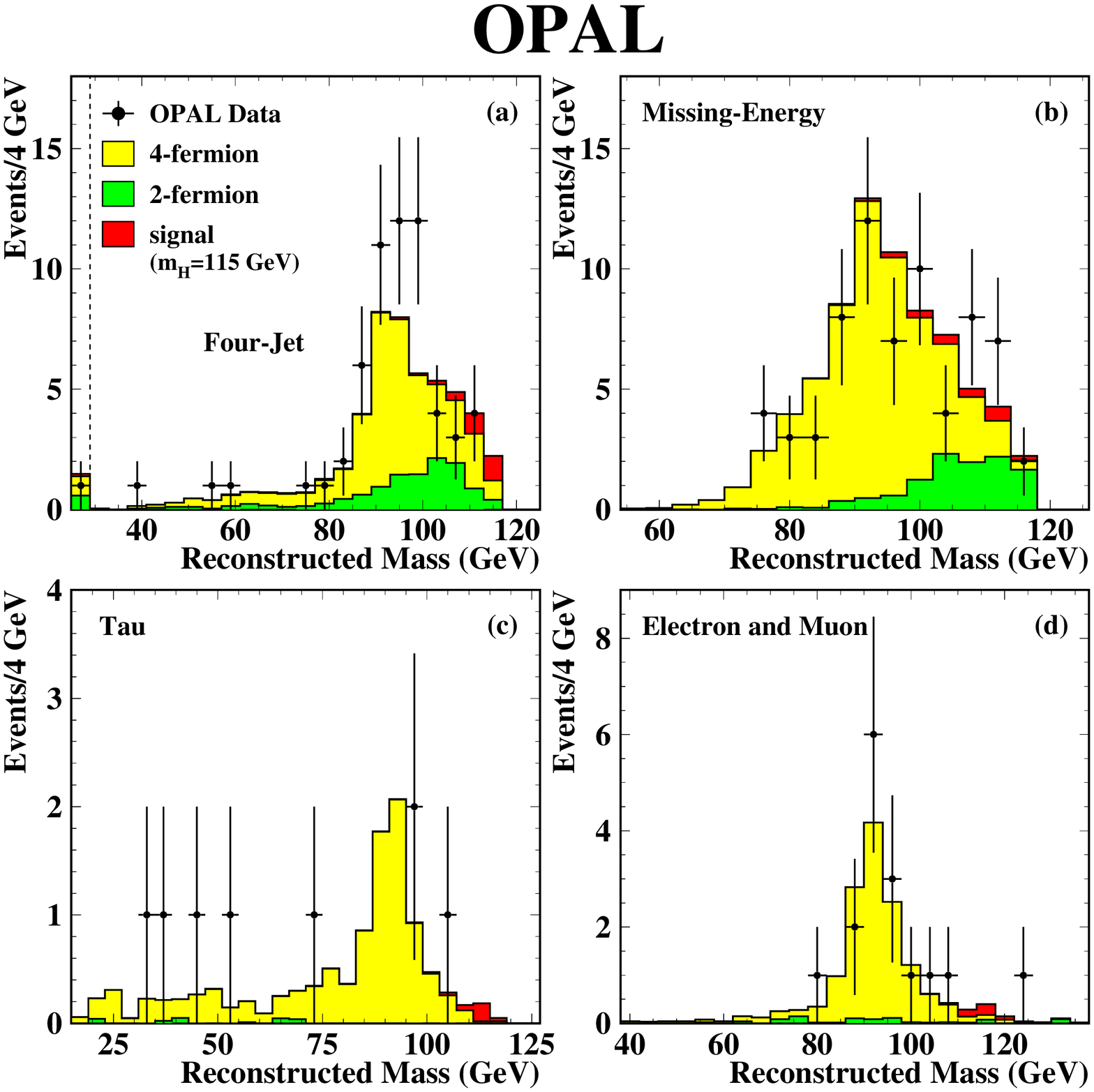,width=\textwidth}
}
\vspace*{-0.3cm}
\caption[]{\label{fig:sm2000loose}\sl 
The reconstructed mass distribution for the selected events in the 1999 and 2000 data
for (a) the four-jet channel, 
(b) the missing-energy channel,
(c) the tau channels, 
and (d) the electron and muon channels combined.
The first bin in (a) contains all
events with $\chi^2$ probability of the
$\mathrm{H^{0}Z^{0}}$ 5C kinematic fit $< 10^{-5}$ for chosen jet-pairings.
The dark (light) grey area shows the expected contribution from the
$\qq (\gamma)$ (four-fermion) process. The \SM\ signal expectation for 115 GeV is
shown with the very dark histograms on top of the \SM\ backgrounds.
}
\end{figure}

The distributions of the reconstructed masses 
of the selected events are shown 
in Figure~\ref{fig:sm2000loose}.
Note that all data taken in the wide $\sqrt{s}$ range from 192 to 209~GeV
are summed in the figure, while the expected 
signal rates strongly depend on $\sqrt{s}$. 
The method used to optimise the sensitivity
to the signal is described in Section~\ref{sec:clmethod}.
The background accumulation at high reconstructed
mass in the four-jet channel
is dominated by \qq\ events and jet-pairing combinatorial backgrounds
from the \Zo\Zo\ process.
The \qq\ background in the missing-energy channel also clusters 
at high reconstructed mass because \qq\ events 
passing the selection requirements
are largely composed of events with two or more undetected,
energetic initial state radiation
photons with a small momentum
sum along the beam direction.  
The jets in such events are nearly back-to-back, 
which results in values of $m^{\mathrm{rec}}_{\mathrm{H}}$ near the maximum
kinematically allowed in the fits, $\sqrt{s}-\mZ$.

\subsection{Confidence Level Calculations}
\label{sec:clmethod}

After the event selections, 
all results from the various search channels 
are combined  to test for the presence of a \SM\ Higgs boson signal.
Previous data taken at centre-of-mass energies near or below 189~GeV
have a negligible impact on the sensitivity to Higgs boson signals
with $\mH>100$~GeV and are not included.
The cross-sections used in computing the confidence level (CL)
include the effects of $\WW$ and $\Zo\Zo$ fusion processes 
and their interference with the \Ho\Zo\ process
in the missing-energy and electron channels respectively, 
as calculated using HZHA3~\cite{HZHA3}.

In order to compute the confidence levels, a test statistic is
defined which expresses how signal-like the data are.  
The confidence levels are computed from the test statistic of the
observed data and the expected distributions of the test statistic
in a large number of simulated experiments under two hypotheses:
the background-only hypothesis and the signal+background hypothesis.
The test statistic chosen is the likelihood ratio $Q$, the ratio of
the probability of observing the data given the signal+background hypothesis
to the probability of observing the data given the background-only
hypothesis~\cite{bib:kendallstuartord,LEPHIGGS202}.
The results of all search channels are expressed in fine bins of
discriminating variables, such as $m^{\mathrm{rec}}_{\mathrm{H}}$.
The expected signal strength depends strongly on $\sqrt{s}$, hence
the results are considered separately in fine divisions of $\sqrt{s}$.
In each bin of each channel at each $\sqrts$ the expected Higgs boson signal,
$s_i$, and the Standard Model background rate, $b_i$, are estimated,
and the observed data counts, $n_i$, are reported.
The $s_i$ depend on the mass of the Higgs boson under study (the
``test mass'').   Each bin is considered to be a statistically independent
search obeying Poisson statistics.  The likelihood ratio
$Q$ can then be computed~\cite{LEPHIGGS202} as
$\ln Q=-\sum_i s_i +\sum_i n_i\ln(1+s_i/b_i)$.
Each event has a weight in the sum which depends on the 
signal-to-background ratio in the bin in which it is found; events may
be classified by their local $s/b$ values.
The confidence level for the background hypothesis is
($1-{\mathrm{CL}}_b$)~\cite{LEPHIGGS202}
which is the probability in an ensemble of background-only experiments
of observing a more signal-like $Q$ than is
actually observed: $1-{\mathrm{CL}}_b=P(Q>Q_{\mathrm{obs}}|{\mathrm{background}})$.
A low value of $(1-{\mathrm{CL}}_b)$ indicates an excess of candidates in data 
compared to the expectation from background.  The distribution of
($1-{\mathrm{CL}}_b$) is uniform between~0 and~1 in an ensemble
of background-only experiments.

Similarly, the confidence level for the signal+background hypothesis is
${\mathrm{CL}}_{s+b}=P(Q\le Q_{\mathrm{obs}}|{\mathrm{signal+background}})$,
and is used to exclude the signal+background hypothesis if it has
a small value.
To eliminate the possibility of excluding a signal to which there is
no sensitivity, a third quantity is defined 
${\mathrm{CL}}_s={\mathrm{CL}}_{s+b}/{\mathrm{CL}}_b$~\cite{LEPHIGGS202}.
Results are also presented in terms of the signal rate limit 
$n_{95}=g_{\mathrm{min}}\sum_i s_i$, where $g_{\mathrm{min}}$ is the
smallest number such that the signal hypothesis consisting of
$g_{\mathrm{min}}s_i$ in each bin yields ${\mathrm{CL}}_s=0.05$.  The
signal rate limit $n_{95}$ depends on the test mass.
The technique used to perform the computation of the confidence levels is the same
as is used in~\cite{prold}.

\begin{figure}[ht]
\vspace*{-1.1cm}
\centerline{
\epsfig{file=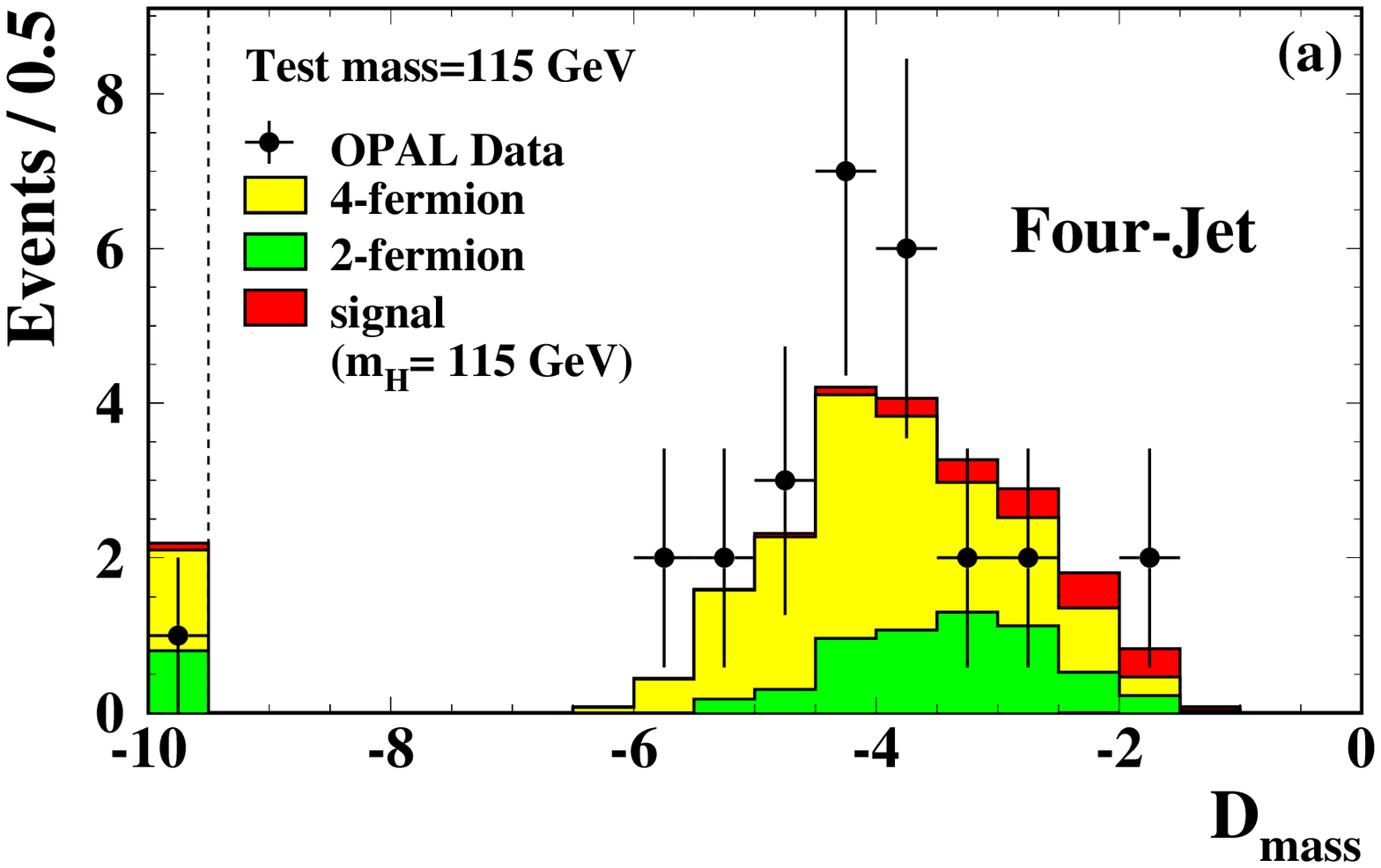,width=0.7\textwidth}
}
\vspace*{-1.5cm}
\centerline{
\epsfig{file=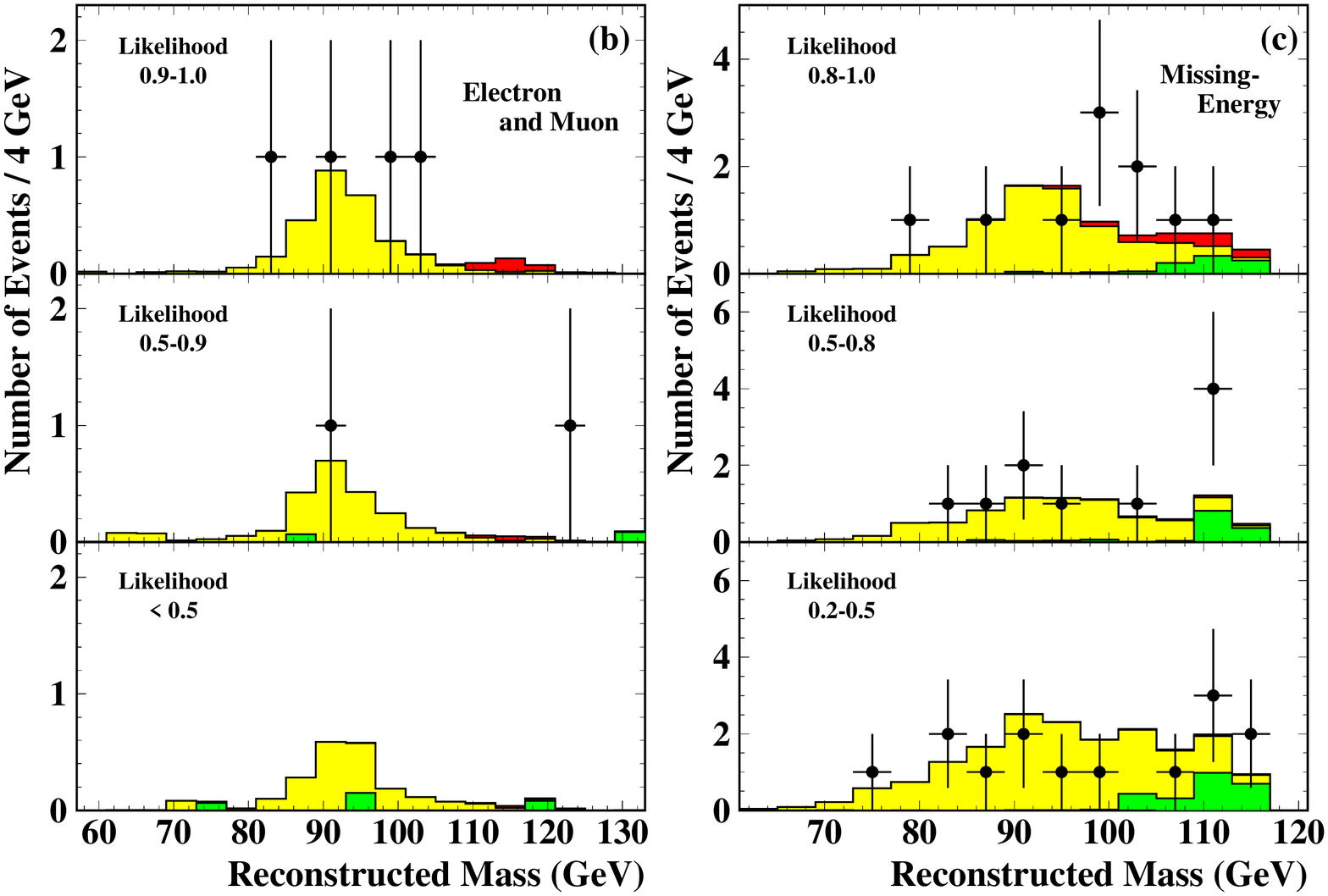,width=1.1\textwidth}
}
\vspace*{-0.2cm}
\caption[]{\label{fig:sm200nem}\sl 
(a) The distribution of 
the variable ${\cal D}_{mass}$ 
in the four-jet channel for a test mass of 115 GeV.
(b) The reconstructed mass distribution in the electron and muon channels for 
the likelihood ranges of 0.9--1.0, 0.5--0.9, and range between
the selection cut values (0.2 for the electron channel
and 0.3 for the muon channel) and 0.5.
(c) The reconstructed mass distribution in the missing-energy channel for 
the likelihood ranges of 0.8--1.0, 0.5--0.8, and 0.2--0.5.
In all distributions, the data for $\sqrt{s}\ge 204$ GeV are shown
with points, and the expected
$\qq(\gamma)$ backgrounds are shown with
dark-shaded histograms and the expected four-fermion backgrounds are
shown with light-shaded histograms.  The expected distributions from 
a 115 GeV Higgs boson signal are shown with very dark-shaded histograms.
}
\end{figure}

In our previous papers, the only discriminating variables used were the values
of $m^{\mathrm{rec}}_{\mathrm{H}}$.  Here, the discriminating power is improved
by combining $m^{\mathrm{rec}}_{\mathrm{H}}$ with other variables.
In the four-jet channel, after the jet pairing assignment
and mass determination,
a new test-mass-dependent variable 
${\cal {D}}_{mass}$ is formed using a likelihood technique 
in order to investigate the compatibility with the signal production hypothesis
for a Higgs boson of a specific test mass.
The variable ${\cal {D}}_{mass}$ is based on the following four quantities:
1) the combined b-tagging variable ${\cal B}_{2jet}^{H}$ 
defined by ${\cal B}_{2jet}^{H} \equiv {\cal {B}}_{1} \cdot {\cal {B}}_{2} / ({\cal {B}}_{1} \cdot {\cal {B}}_{2} + (1- {\cal {B}}_{1}) \cdot (1-{\cal {B}}_{2}))$ 
for the two jets with the most significant b~tags;
2) the energy difference between the most energetic and the least energetic
jets in the event;
3) $\beta_{\rm min}$, a selection likelihood 
variable (see~\cite{prold});
and 4) the reconstructed Higgs boson mass $m^{\mathrm{rec}}_{\mathrm{H}}$.
Signal Monte Carlo samples were generated
with Higgs boson masses in 1 GeV steps, 
and ${\cal {D}}_{mass}$ is interpolated between neighbouring test masses.
The distribution of ${\cal D}_{mass}$ is shown for 
a test mass of 115~GeV in Figure~\ref{fig:sm200nem}(a).

In the missing-energy, electron and muon channels,
the selection likelihood value, ${\cal L}^{\rm H^0Z^0}$, 
is used to form a two-dimensional discriminant 
(${\cal L}^{\rm H^0Z^0}$, $m^{\mathrm{rec}}_{\mathrm{H}}$).
The cut on ${\cal L}^{\rm H^0Z^0}$ 
is chosen to optimise the sensitivity of this new discriminant. 
In Figures~\ref{fig:sm200nem}(b) and (c),
the reconstructed mass distributions are shown in slices of the selection
likelihoods
for the electron and muon channels, and for the missing-energy channel, 
respectively.  In each channel, the enrichment of the signal depends
on both the likelihood value and on $m^{\mathrm{rec}}_{\mathrm{H}}$.
For the tau channels, only the reconstructed mass is used
for the discriminant as in ~\cite{pr189}.

The systematic uncertainties on the signal and background 
expectations in each channel are treated using an 
extension of the method described in~\cite{ref:cousins}. 
Uncertainties described in Section~\ref{sec:syserror} are assumed
to be 100\% correlated if they arise from the same source
in different channels, in the signal and background estimations,
and at different centre-of-mass energies.
The current uncertainty on the beam energy for the 2000 data
is expected to be of the 
order of 100~MeV, and 
would therefore affect the limits by $\sim$~200~MeV.
The uncertainty on the integrated 
luminosity is estimated to be 0.3\%. 
Both of these errors are neglected.

\section{Results}
\label{sec:smcl}

\begin{figure}[htbp]
\centerline{
\epsfig{file=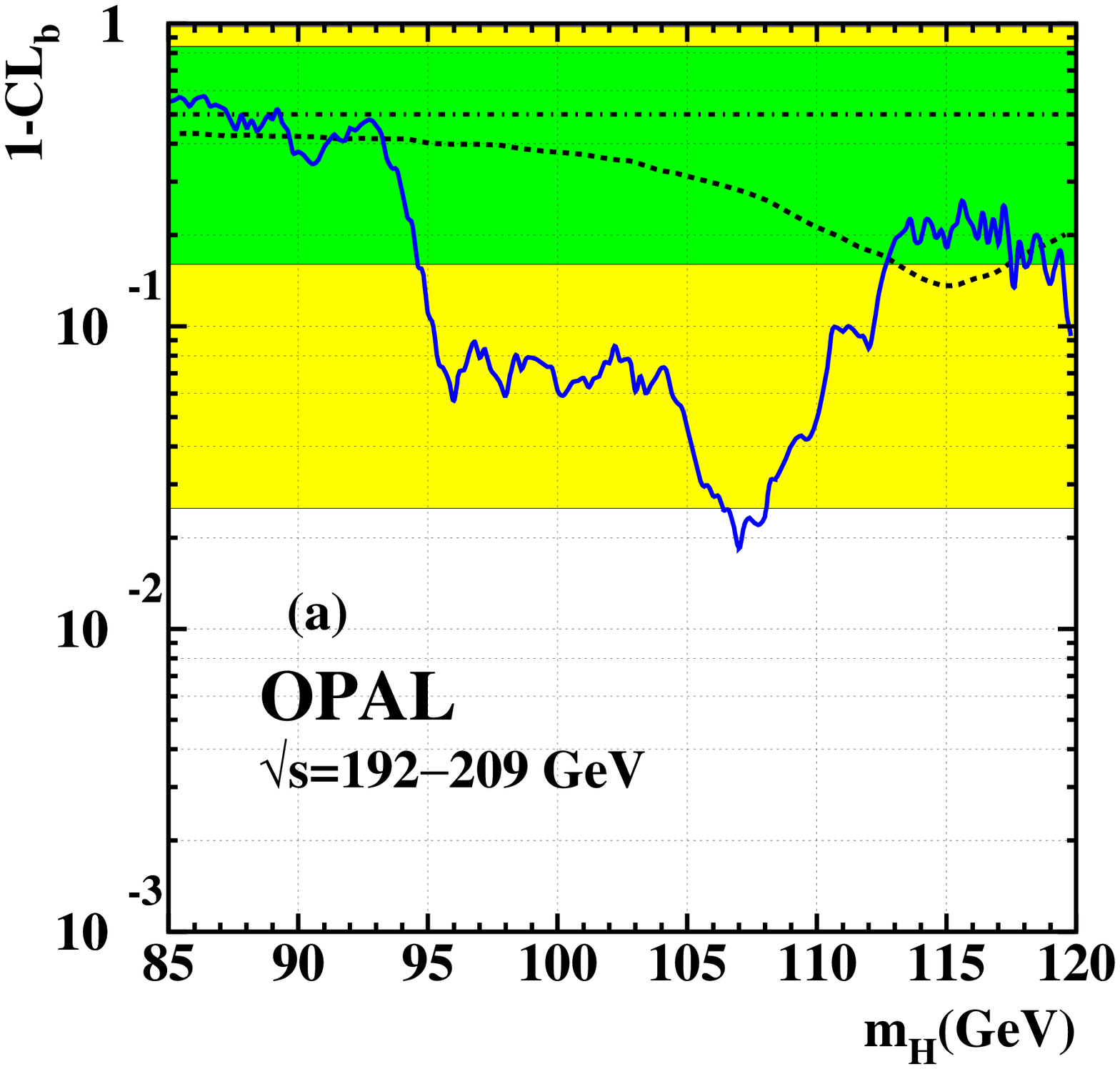,width=0.6\textwidth}\\
\epsfig{file=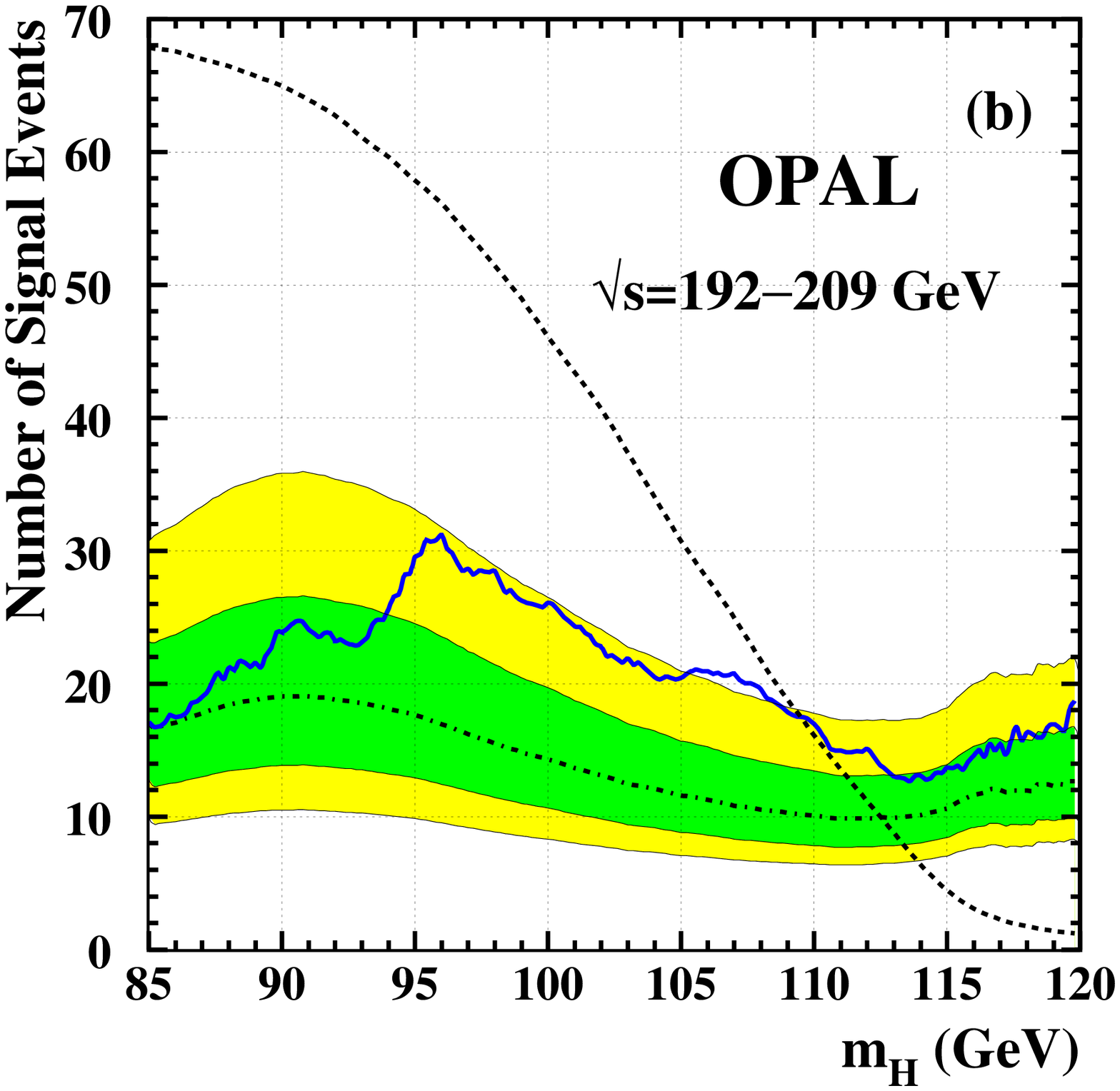,width=0.6\textwidth}
}
\vspace*{-0.2cm}
\caption[]{\label{fig:smclb}\sl 
(a) The confidence level for the background-only hypothesis, 
$(1-{\mathrm{CL}}_b)$, as a function of the Higgs boson test mass.  
The dotted curve represents the median expectation assuming the presence
of the Standard Model Higgs boson with a 115 GeV mass.
The dark (light) shaded bands indicate the 68\% (95\%) probability
intervals centred on 0.5,
the median expectation in the absence of a signal, which is indicated with
a dot-dashed line.
(b) Upper limits on the signal counts
at the 95\% confidence level ($n_{95}$), as observed (solid line) 
and the expected median (dot-dashed line) for background-only experiments,
as a function of the Higgs boson test mass.
The expected rate of the accepted signal counts 
for a Standard Model Higgs boson with a mass equal to the test mass
is shown with the dotted line.
The shaded bands are the 68\% and 95\% probability intervals
centred on the median background
expectation.  The range for the test mass \mH\ is chosen in both (a) and (b)
to extend to 85~GeV to show the consistency of the 1999 and 2000
OPAL data with the background expectations near the \Zo\ peak, even though previous
OPAL search results~\protect\cite{pr189} are sensitive to test masses below 100~GeV.
}
\end{figure}

Figure~\ref{fig:smclb}(a) shows $(1-{\mathrm{CL}}_b)$ as a function of the
test mass \mH.  It attains its lowest value of 0.02 at $\mH=107$~GeV,
indicating a local excess of candidates.  
The probability to observe such an excess anywhere in the range
of test masses between 100 and 120~GeV is approximately 10\%, estimated
from the size of the range and the reconstructed mass resolution.
The value of $(1-{\mathrm{CL}}_b)$ observed at \mH=115~GeV is 0.2.
Figure~\ref{fig:smclb}(a) also shows the expected $(1-{\mathrm{CL}}_b)$ in the 
presence of a 115~GeV Higgs boson signal.

The signal rate limit $n_{95}$ is shown as a function of \mH\ in
Figure~\ref{fig:smclb}~(b) together with its median expectation in
an ensemble of background-only experiments.  Figure
~\ref{fig:smclb}~(b) also shows the expected accepted
signal rate.  Where the signal rate curve crosses the $n_{95}$ curve
is the 95\%~CL exclusion limit on \mH, and the expected limit
is where the median expected $n_{95}$ curve crosses the accepted signal
rate curve.  A lower mass bound of 109.7~GeV is obtained, and
the expected limit is 112.5~GeV.  In particular, the hypothesis
\mH=107~GeV is excluded at the 98\% CL (CL$_s=0.02$)
even in the presence of the excess candidates
because the excess in the data is not large enough to be consistent with
the expected signal rate from a Standard Model Higgs boson
of that mass.

\begin{figure}[ht]
\vspace*{-0.5cm}
\centerline{
\epsfig{file=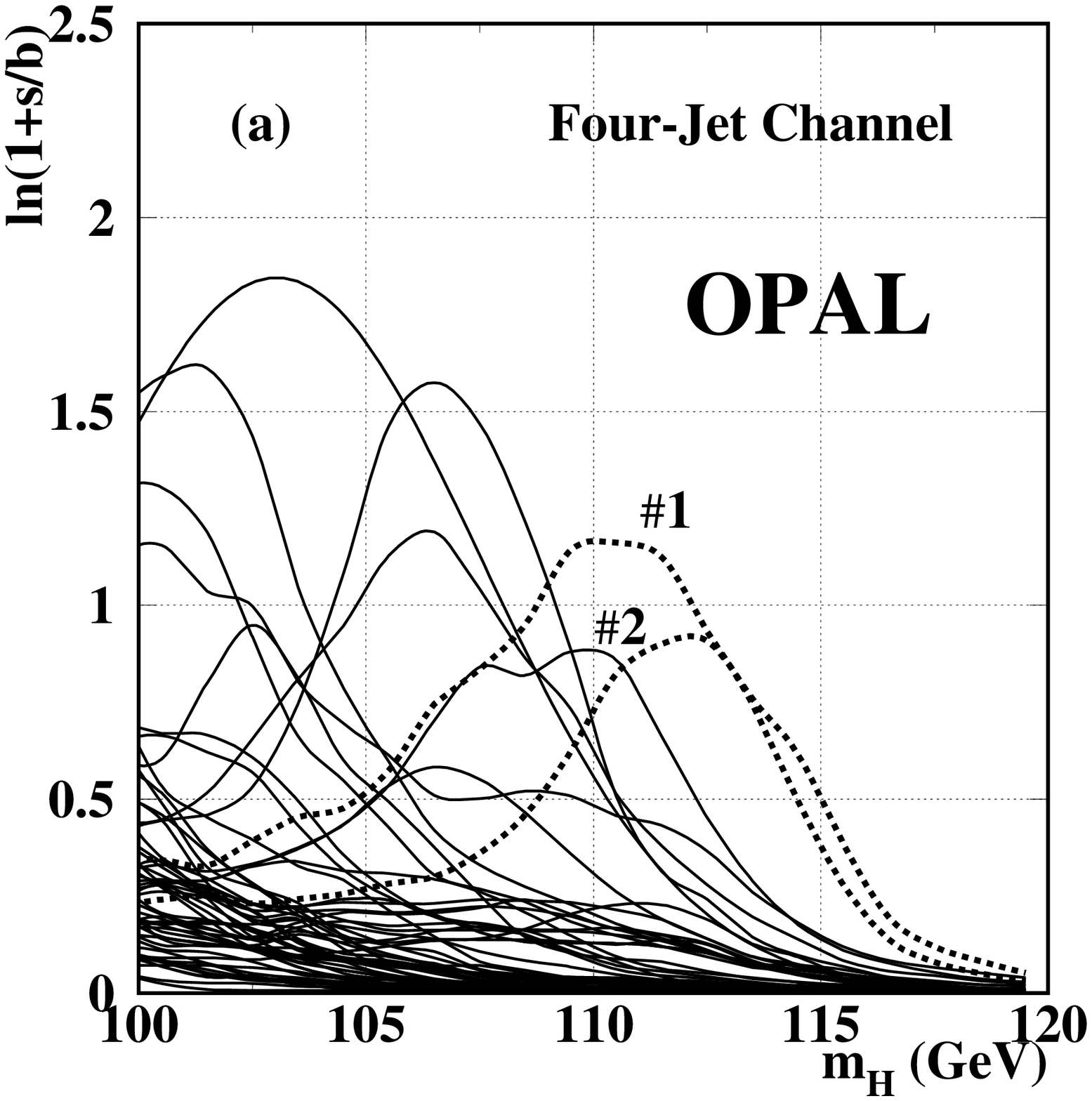,width=0.55\textwidth}\epsfig{file=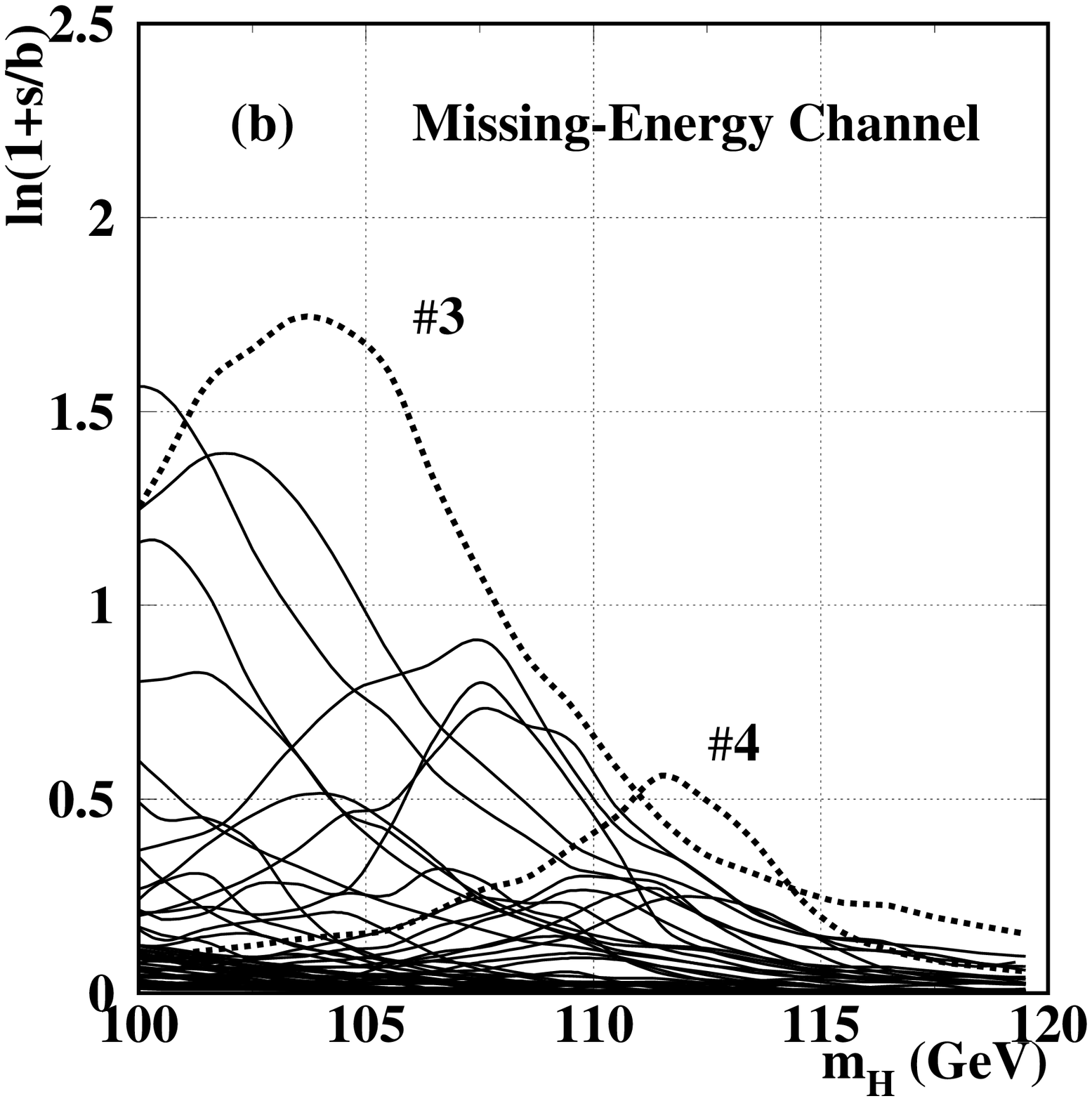,width=0.55\textwidth}
}
\vspace*{-0.5cm}
\centerline{
\epsfig{file=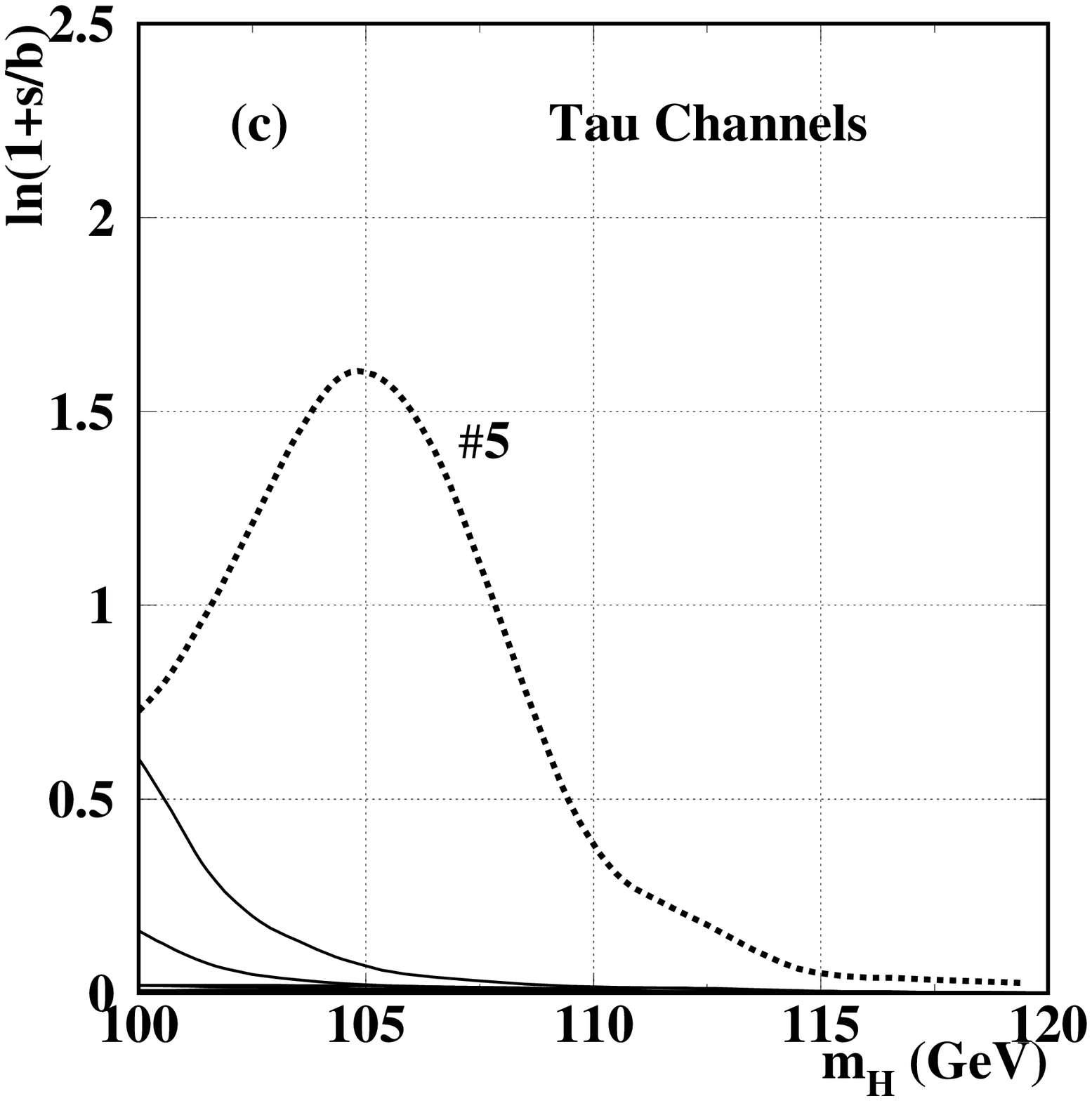,width=0.55\textwidth}\epsfig{file=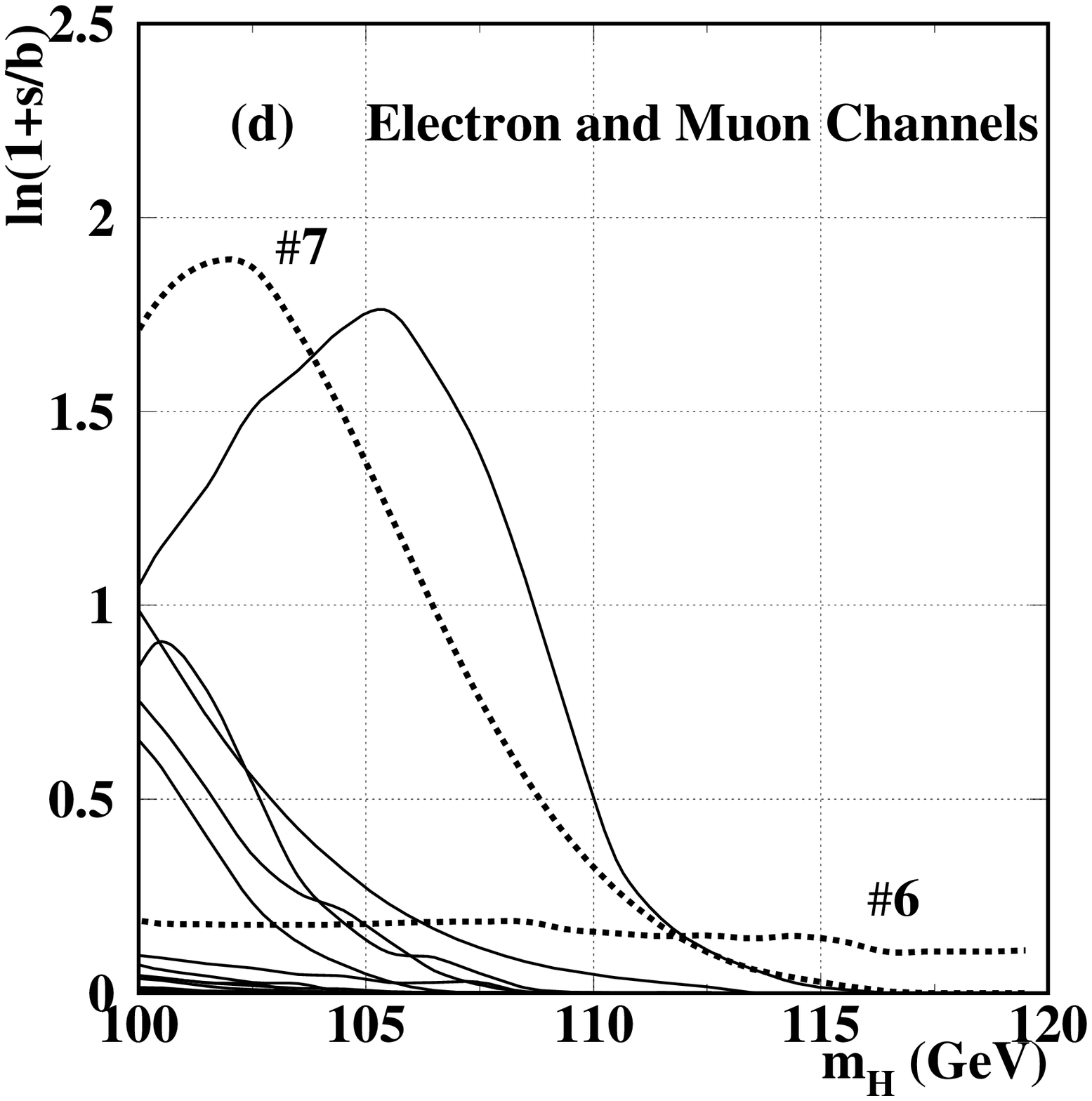,width=0.55\textwidth}
}
\vspace*{-0.1cm}
\caption[]{\label{fig:sm_evolution}\sl 
The $\ln(1+s/b)$ distributions as a function of the Higgs boson test mass 
for each candidate collected in 1999 and 2000 for
(a) the four-jet channel, (b) the missing-energy channel, (c) the
tau channels, and (d) the electron and muon channels.
Each curve in the plots represents the $\ln(1+s/b)$ of each candidate 
as a function of the Higgs boson test mass. 
The dotted lines 
show the contributions of the candidates listed in Table~\ref{tab:4j-cand}.
}
\end{figure}

The candidates with the largest 
weights, $\ln (1+s/b)$, in each channel 
for test masses of 115~GeV are listed in 
Table~\ref{tab:4j-cand}. 
Figure~\ref{fig:sm_evolution} shows 
the distributions of $\ln (1+s/b)$ of each candidate
as a function of the Higgs boson test mass for the
candidates collected in 1999 and 2000.  The region of test mass for which
a candidate's contribution is significant depends on the mass resolution
of the candidates in the channel, and the normalization of the curve depends
on the candidate b-tags, the kinematic variables, the Higgs boson cross-section
and the local background near the reconstructed candidate mass.
Deviations from smoothness of the curves are due to
Monte Carlo statistics; this uncertainty is included in the
confidence level computation.

The most significant candidate for a Higgs boson search for
a test mass $\mH=115$~GeV
(candidate \#1) is found in the four-jet channel.
For the jet-pairing chosen by the jet-pairing likelihood function, the event
has a reconstructed Higgs boson mass
of 110.7~GeV.   The second most significant candidate (\#2) is also
found in the four-jet channel, with a reconstructed mass of 112.6~GeV.
No jet pairing yields a reconstructed Higgs boson mass close to
the $\Zo$ mass in either of these two candidates.

The observed low value of 
$(1-{\mathrm{CL}}_b)$ at 107 GeV is caused by candidates
which have relatively high weights 
at around 105--110 GeV seen in Figure~\ref{fig:sm_evolution}.
For the tau channel candidate \#5,
the reconstructed mass is taken from the invariant
mass of the jets after a 3C kinematic fit 
where the tau pair mass is constrained to the Z$^0$ mass.
The tau pair mass is 91~GeV if a 2C fit~\cite{pr189}
is performed.
The muon channel candidate \#7 significantly affects
the results of the confidence level calculations around 100--105 GeV
since its likelihood ${\cal L}^{\rm H^0Z^0}$
is very close to one. 

\begin{figure}[ht]
\vspace*{-1cm}
\centerline{
\epsfig{file=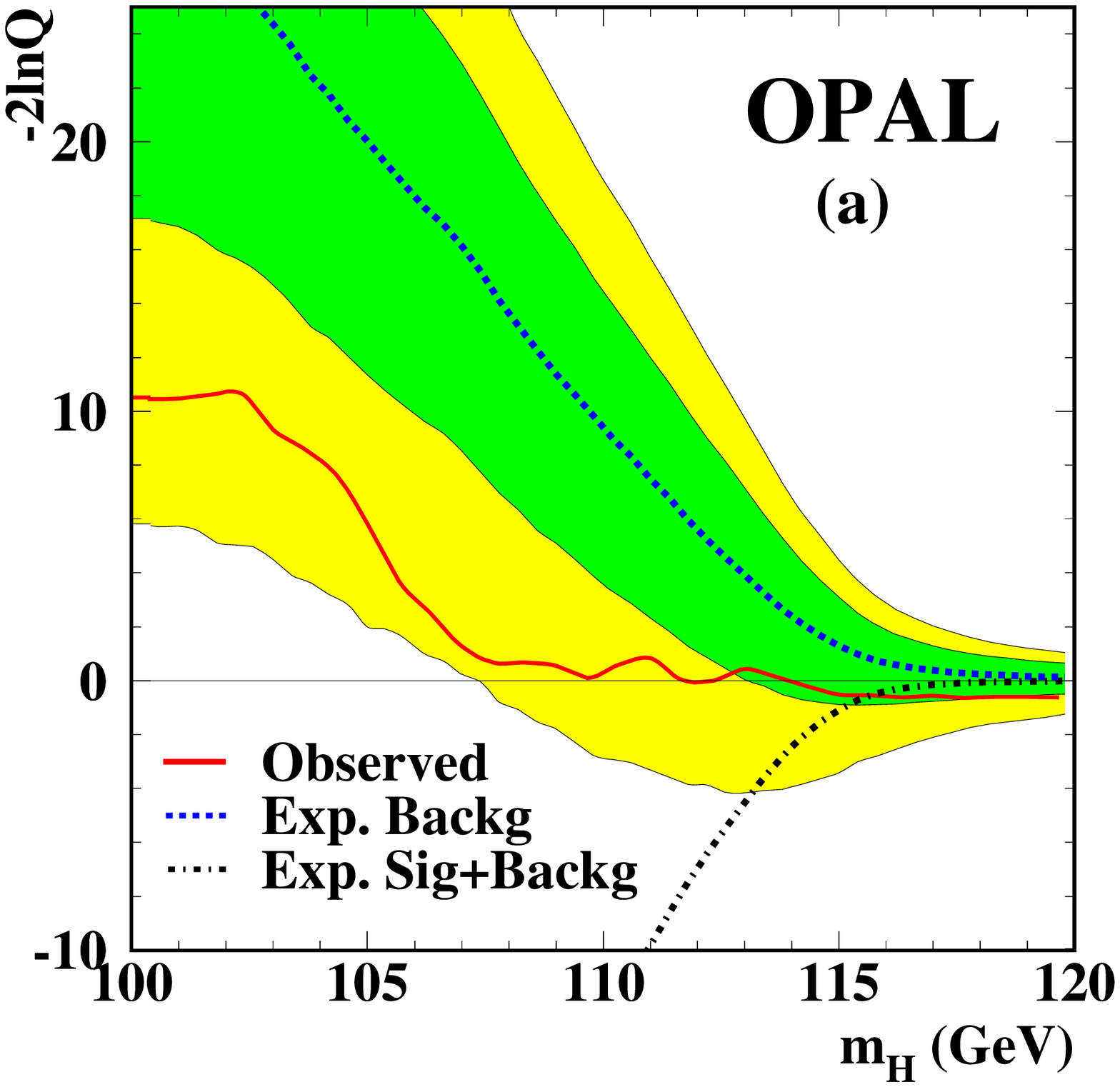,width=0.53\textwidth}\epsfig{file=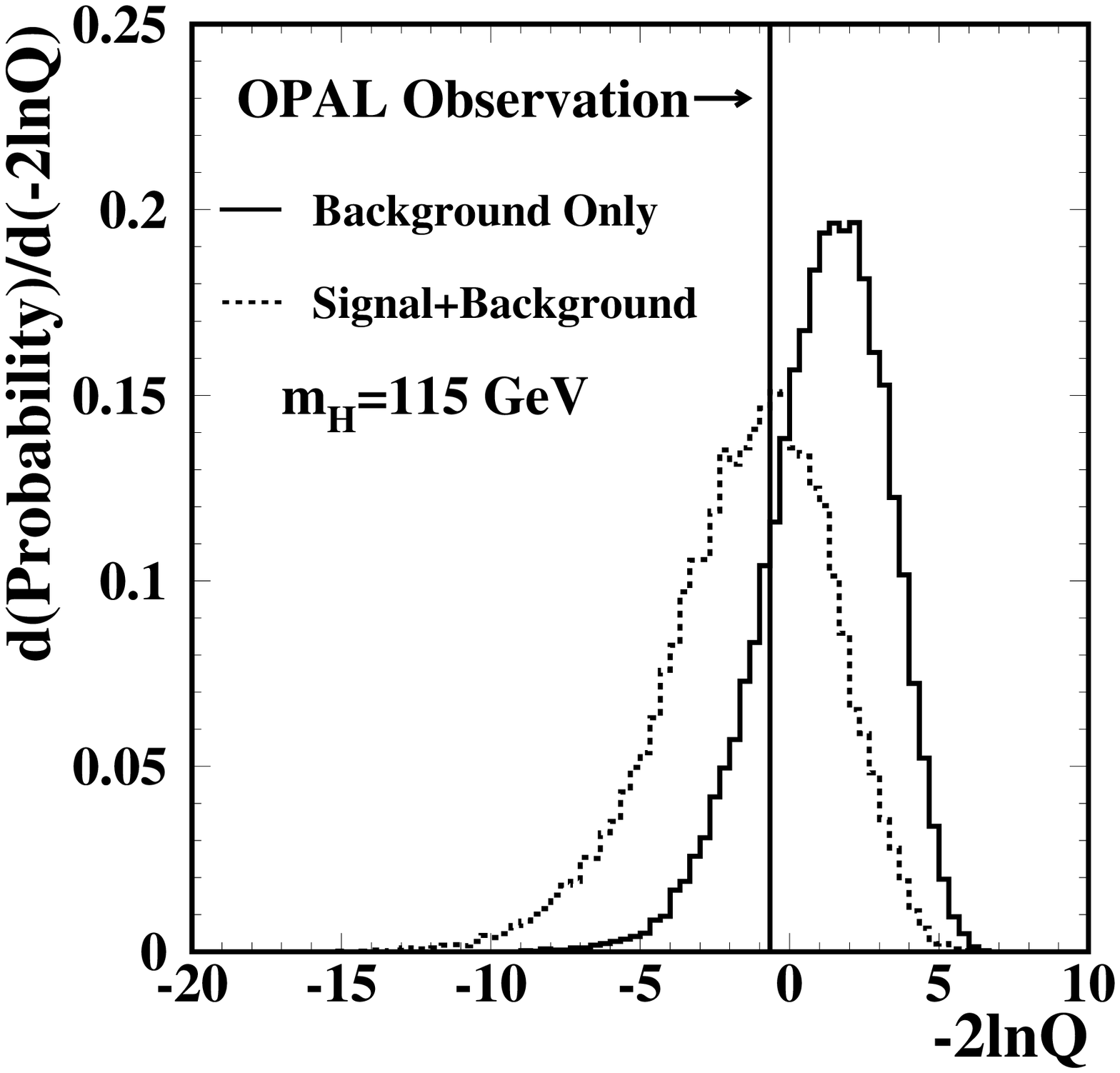,width=0.6\textwidth}
}
\vspace*{-0.1cm}
\caption{\label{fig:minus2lnq}\sl
(a) The log-likelihood ratio $-2\ln Q$ 
comparing the relative consistency of the data
with the signal+background hypothesis and the 
background-only hypothesis, as a function of the test mass \mH. 
The observation for the data 
is shown with a solid line. 
The dashed line indicates the median background expectation and
the dark (light) shaded band shows the 68\% (95\%) probability
intervals centred on the median.
The median expectation in the presence of a signal
is shown with a dot-dashed line where the hypothesized signal mass is
the test mass.
(b) The $-2 \ln Q$ distribution expected in 
a large number of fictitious background-only 
experiments (solid histogram), and in a large number of
fictitious experiments in the presence of a 115~GeV Higgs
boson (dashed histogram).  The observation
in the data is shown with a vertical solid line.
}
\end{figure}

The observed $-2\ln{Q}$
is shown as a function of the test mass \mH\ in
Figure~\ref{fig:minus2lnq}~(a).
Also shown are the 68\% and 95\% probability contours
centred on the median expectation.  Figure~\ref{fig:minus2lnq}~(b)
shows the probability density functions of $-2\ln Q$ for the
signal+background hypothesis with $\mH=115$~GeV, and also for the
background hypothesis.  The separation between the
two hypotheses is not strong.  The background confidence
level ($1-{\mathrm{CL}}_b$)=0.2 is
the integral of the background-only probability density to the left
of the data observation, and ${\mathrm{CL}}_{s+b}=0.4$ is the integral
to the right of the data observation of the signal+background curve.
The data are slightly more
consistent with the presence of a 115 GeV Higgs boson than with the
background alone.

\section{Conclusions} 

A search for the Standard Model Higgs boson has been performed 
with the OPAL detector at LEP based on the full data sample
collected at $\sqrt{s}\approx$192--209~GeV in 1999 and 2000.
The largest deviation with respect to the expected \SM\ background
in the confidence level for the background hypothesis, $(1-{\mathrm{CL}}_b)$, 
is observed for a Higgs boson mass of 107 GeV with a 
minimum $(1-{\mathrm{CL}}_b)$ of 0.02, 
but the observed excess is less
significant than is expected for a Standard Model Higgs boson 
with a 107 GeV mass.
A lower bound of 109.7~GeV on the mass of the \SM\ Higgs boson is obtained
at the 95\% confidence level while the median expectation for
the background-only hypothesis is 112.5 GeV.
For a Higgs boson with a mass of 115~GeV,
$(1-{\mathrm{CL}}_b)$ is approximately 0.2 while
CL$_{s+b}$ is approximately 0.4, indicating that the data slightly
favour the hypothesis that a signal is present, but also that the
data are consistent with the background hypothesis.
These data alone provide little discrimination
between the signal+background and background hypotheses
for Higgs boson masses above 112~GeV, but more statistically powerful conclusions may
be reached by combining the data presented here with those of the other
LEP experiments~\cite{LEPHIGGS00COMB}.

\section*{Acknowledgements}

We particularly wish to thank the SL Division for the efficient operation
of the LEP accelerator at all energies
 and for their continuing close cooperation with
our experimental group.  We thank our colleagues from CEA, DAPNIA/SPP,
CE-Saclay for their efforts over the years on the time-of-flight and trigger
systems which we continue to use.  In addition to the support staff at our own
institutions we are pleased to acknowledge the  \\
Department of Energy, USA, \\
National Science Foundation, USA, \\
Particle Physics and Astronomy Research Council, UK, \\
Natural Sciences and Engineering Research Council, Canada, \\
Israel Science Foundation, administered by the Israel
Academy of Science and Humanities, \\
Minerva Gesellschaft, \\
Benoziyo Center for High Energy Physics,\\
Japanese Ministry of Education, Science and Culture (the
Monbusho) and a grant under the Monbusho International
Science Research Program,\\
Japanese Society for the Promotion of Science (JSPS),\\
German Israeli Bi-national Science Foundation (GIF), \\
Bundesministerium f\"ur Bildung und Forschung, Germany, \\
National Research Council of Canada, \\
Research Corporation, USA,\\
Hungarian Foundation for Scientific Research, OTKA T-029328, 
T023793 and OTKA F-023259.



\newpage

\begin{table}[tb]
\begin{center}
{\footnotesize
\begin{tabular}{|c||c||r||r||r|r||c|c|}\hline
Channel & Cut & Data & Total bkg. & q\=q($\gamma$) & 4-fermi. & 
\multicolumn{2}{|c|}{Eff. [\%] (signal events)}\\ \cline{7-8}
 &  &  &  &  &  & 110~GeV & 115~GeV \\\hline\hline
Four-jet &Presel.&  3820& 3609.9& 760.8& 2859.4& 85.2 & 86.9 \\
\cline{2-8}
&${\cal L}^{\rm HZ}$ &
      60&    49.8$\pm$6.0 &  12.8&   37.0&45.5 (9.04$\pm$0.41)& 41.8 (2.24$\pm$0.10)\\
\hline\hline
Missing-E & Presel. & 354& 334.5&  57.0&  277.5 & 56.1 & 49.5\\
\cline{2-8}
&${\cal L}^{\rm HZ}$ &
     68&    69.7$\pm$8.6&  10.6&   59.1&50.0 (4.70$\pm$0.28)& 43.9 (1.68$\pm$0.10)\\
\hline\hline
Tau&Presel. &  343&   334.5&  57.0&  277.5&48.3&43.3\\
\cline{2-8}
&Final $\mathcal{L}$ &
    8&    11.1$\pm$1.2&   0.4&   10.7&29.5 (0.98$\pm$0.05)& 22.9 (0.25$\pm$0.01)\\
\hline\hline
Electron&Presel.&  429&   378.6& 171.0&  207.6&72.7&71.3\\
\cline{2-8}
&${\cal L}^{\rm HZ}$ &
   6&     8.5$\pm$1.3&   0.3&    8.2&52.9 (0.66$\pm$0.02)&48.7 (0.17$\pm$0.004)\\
\hline\hline
Muon&Presel.&  79&    66.2&  36.8&   29.4& 70.3&70.9\\
\cline{2-8}
&${\cal L}^{\rm HZ}$ &
   10&     7.0$\pm$1.0&   0.2&    6.8&59.1 (0.80$\pm$0.02)& 59.9 (0.23$\pm$0.006)\\
\hline
\end{tabular}
}
\caption[]{\label{tab:tab1new}\sl 
  The number of events after preselection 
  and after the final likelihood selection for 
  the 192--209 GeV data and the expected background.
  The errors on the total background and the expected Higgs signal 
  include all systematic errors.
  The last two columns show the detection efficiencies 
  (and the numbers of expected signal events in parentheses) 
  for a Higgs boson with \mH=110~GeV and 115 GeV.  
  For the four-jet channel, the efficiency is computed only
  for \Ho\ra~\bb\ decays, and for the tau channel 
  for the processes \Ho\Zo\ra\tautau(\Ho\ra all) 
  or \Ho\Zo\ra\qq\tautau\ assuming \SM\ branching fractions.
  For other channels, 
  the efficiency is for all decays of the \SM\ Higgs boson.
}
\end{center}
\end{table}

\begin{table}[htb]
\begin{center}
\begin{footnotesize}
\begin{tabular}{|c|c|c|c|c|c|c|c|}\hline
\multicolumn{8}{|c|}{Candidates} \\\hline
Candidate & Channel & $m{\mathrm{_H^{rec}}}$ & ${\cal L}^{\rm H^0Z^0}$  & 
$E_{\mathrm{CM}}$ & 
\multicolumn{3}{|c|}{ $s/b$ for $\mH$ (GeV)}\\ \cline{6-8} 
         &       & (GeV)                &   & 
(GeV)             & 
105 & 110 & 115 \\\hline
\# 1 ~~&~~Four-jet  ~~&~~ 110.7  ~~&~~ 0.995 ~~&~~206.6~~&~~ 0.64 ~~&~~ 2.09 ~~&~~ 0.70 ~~\\
\# 2 ~~&~~Four-jet  ~~&~~ 112.6  ~~&~~ 0.999 ~~&~~205.4~~&~~ 0.28 ~~&~~ 1.18 ~~&~~ 0.49 ~~\\
\# 3 ~~&~~Missing-Energy ~~&~~ 104.0  ~~&~~ 0.999 ~~&~~205.4~~&~~ 4.55 ~~&~~ 0.96 ~~&~~ 0.28 ~~\\
\# 4 ~~&~~Missing-Energy ~~&~~ 112.1  ~~&~~ 0.853 ~~&~~206.4~~&~~ 0.15 ~~&~~ 0.44 ~~&~~ 0.23 ~~\\
\hline\hline
\# 5 ~~&~~Tau       ~~&~~ 105.3  ~~&~~ 0.993 ~~&~~205.3~~&~~ 4.00 ~~&~~ 0.46 ~~&~~ 0.05 ~~\\
\# 6 ~~&~~Electron  ~~&~~ 124.7  ~~&~~ 0.873 ~~&~~205.4~~&~~ 0.20 ~~&~~ 0.17 ~~&~~ 0.16 ~~\\
\# 7 ~~&~~Muon      ~~&~~ 102.2  ~~&~~ 0.999 ~~&~~205.4~~&~~ 3.02 ~~&~~ 0.38 ~~&~~ 0.04 ~~\\
\hline
\end{tabular}
\end{footnotesize}
\caption{\label{tab:4j-cand}\sl 
The candidates with the largest weights for the 115 GeV Higgs
hypothesis in each channel.
The columns labeled $m{\mathrm{_H^{rec}}}$ and ${\cal L}^{\rm H^0Z^0}$ are 
the reconstructed Higgs mass and selection likelihood values, respectively.
The next three columns give the signal to background ratio (s/b) using the
discriminants which are used in the confidence level calculation. 
The s/b values for Higgs boson test masses of 105, 110 and 115 GeV 
are shown.  
For the four-jet and the missing-energy channels, all candidates
with s/b larger than 0.20 for 115 GeV are listed. For the tau, 
electron and muon channels, the candidates with the 
largest s/b for
the 115 GeV hypothesis are listed in the lower portion of the table.
}
\end{center}
\end{table}

\newpage

\newpage

\newpage

\newpage

\newpage

\newpage

\newpage

\begin{thebibliography}{99}
%
\bibitem{sm}
S.~Glashow, Nucl.~Phys.~{\bf 22} (1961) 579;\\
S.~Weinberg, Phys.~Rev.~Lett.~{\bf 19} (1967) 1264;\\
A.~Salam, ed. N.~Svartholm, {\it Elementary Particle Theory}, 
Almquist and Wiksells, Stockholm (1968) 367.
%
\bibitem{higgs}
P.W.~Higgs, Phys.~Lett.~{\bf 12} (1964) 132;\\
F.~Englert and R.~Brout, Phys.~Rev.~Lett.~{\bf 13} (1964) 321;\\
G.S.~Guralnik, C.R.~Hagen, and T.W.B.~Kibble,
Phys.~Rev.~Lett.~{\bf 13} (1964) 585.
%
\bibitem{higgs_piktalk} 
P.~Igo-Kemenes, presentation given to
the LEP Experiments Committee open session, 
\mbox{\tt http://lephiggs.web.cern.ch/LEPHIGGS/talks/index.html},
3~November 2000.
%
\bibitem{higgs_otherlep}
ALEPH Collaboration, R.~Barate \etal, Phys.~Lett.~{\bf B495} (2000) 1;\\  
L3 Collaboration, M. Acciarri \etal, Phys.~Lett.~{\bf B495} (2000) 18.
\bibitem{pr189} 
OPAL Collaboration, G.~Abbiendi \etal, Eur.~Phys.~J.~{\bf C12} (2000) 567.
%
\bibitem{prold} 
OPAL Collaboration, G. Abbiendi \etal, Eur. Phys. J. {\bf C7} (1999) 407;\\
OPAL Collaboration, K. Ackerstaff \etal, Eur. Phys. J. {\bf C1} (1998) 425.
%
\bibitem{detector}
OPAL Collaboration, K.~Ahmet \etal, Nucl. Instr. and Meth. 
{\bf A305} (1991) 275;\\
S. Anderson {\it et al.}, Nucl. Instr. and Meth. A403 (1998) 326;\\
B.E.~Anderson {\it et al.}, IEEE Trans. on Nucl. Science 41 (1994) 845;\\
G.~Aguillion {\it et al.}, Nucl. Instr. and Meth. A417 (1998) 266.
%
\bibitem{HZHA3} 
P.~Janot \etal, in {\em Physics at LEP2,} edited by G.~Altarelli, 
T.~Sj\"ostrand and F.~Zwirner, CERN 96-01 Vol.~2, 309. \\
For HZHA3 and HZHA2, 
see {\em http://alephwww.cern.ch/$\sim$janot/Generators.html}.
%
\bibitem{CEEX}
S.~Jadach, B.F.~Ward and Z.~Was, Phys.~Lett.~{\bf B449} (1999) 97.
%
\bibitem{grc4f}
J. Fujimoto \etal, Comp. Phys. Comm. {\bf 100} (1997) 128;\\
{\it Physics at LEP2}, J.~Fujimoto \etal, CERN 96-01, Vol.~2, 30.
%
\bibitem{pythia}
T. Sj\"ostrand, Comp. Phys. Comm. {\bf 82} (1994) 74;\\
T. Sj\"ostrand \etal, {\em High-Energy-Physics Event Generation
with PYTHIA 6.1}, hep-ph/0010017 (2000), to be published in
Comp. Phys. Comm.
%
\bibitem{opaltune}
OPAL Collaboration, G. Alexander \etal, Z. Phys {\bf C69} (1996) 543.
%
\bibitem{gopal}
J. Allison \etal, Nucl. Instr. and Meth. {\bf A317} (1992) 47.
%
\bibitem{ref:wwxs} 
OPAL Collaboration, G.~Abbiendi \etal, Phys.~Lett.~{\bf B493} (2000) 249.
%
\bibitem{koralW} 
M.~Skrzypek \etal, \CPC{94} (1996) 216;\\
M.~Skrzypek \etal, \PhysLettB{372} (1996) 289. 
%
\bibitem{ref:4funcertainty}  
M.W.~Gr\"unewald and G.~Passarino \etal,
{\it Four-Fermion Production in Electron-Positron Collisions}, 
{\tt hep-ph/0005309} (2000).
%
\bibitem{LEPEWWG} ALEPH, DELPHI, L3, OPAL Collaborations and the LEP Electroweak
Working Group and the SLD Heavy Flavour and Electroweak Groups, 
{\it  A Combination of Preliminary Electroweak Measurements and Constraints
on the Standard Model}, CERN-EP/2000-016 (2000), and references therein.
%
%
\bibitem{gqq-spl}
E. Norrbin and T. Sj\"ostrand, 
{\em QCD Radiation off Heavy Particles}, 
{\tt hep-ph/0010012} (2000).
\bibitem{WWpaper}
OPAL Collaboration, G.~Abbiendi \etal, 
{\em Measurement of the Mass and Width of the W Boson in 
${\mathrm{e^+e^-}}$ Collisions at 189 GeV}, 
CERN-EP-2000-099, submitted to Phys. Lett. B.
%
\bibitem{bib:kendallstuartord} 
A.~Stuart and J.~K.~Ord, 
{\it Kendall's Advanced Theory of Statistics}, Vol.~2, Ch.~23
5th Ed., Oxford University Press, New York, (1991).
%
\bibitem{LEPHIGGS202} 
ALEPH, DELPHI, L3, OPAL Collaborations, and
the LEP working group for Higgs boson searches,
{\em Search for Higgs bosons: Preliminary combined results using LEP data 
collected at energies up to 202 GeV,} 
CERN-EP/2000-055 (2000)
%
%
\bibitem{ref:cousins} 
R.D.~Cousins and V.L.~Highland, Nucl.~Instr.~and~Meth. {\bf A320} (1992) 331.
%
\bibitem{LEPHIGGS00COMB} ALEPH, DELPHI, L3 and OPAL Collaborations, and the
LEP Higgs Working Group, in preparation.
%
\end{thebibliography}
\end{document}